\documentclass[epjST,final]{svjour}

\usepackage{graphics}
\usepackage{amsmath}

\begin{document}

\title{Visibility and stability of superstripes in a spin-orbit-coupled Bose-Einstein condensate}
\author{Giovanni I. Martone\thanks{\email{martone@science.unitn.it}}}
\institute{INO-CNR BEC Center and Dipartimento di Fisica, Universit\`{a} di Trento, 38123 Povo, Italy}

\abstract{
We consider a spin-$1/2$ Bose-Einstein condensate with equal Rashba and
Dresselhaus spin-orbit coupling. After reviewing some relevant features of the quantum phases
of the system, we present a short study on how their properties are changed
by the presence of non-zero magnetic detunings and spin-asymmetric interactions.
At small values of the Raman coupling and of the magnetic field the so-called stripe phase
occurs, which displays both superfluidity and periodic density modulations, in analogy with
supersolids. We finally review a recent proposal (Phys. Rev. A \textbf{90}, 041604)
to improve the visibility of the fringes, based on the space
separation of the two spin components into a 2D bi-layer configuration and on the application
of a $\pi/2$ Bragg pulse, and we show that this new configuration also yields a sizable increase
of the stability of the stripe phase against magnetic fluctuations.
}

\maketitle

\section{Introduction}
\label{sec:intro}
The recent experimental realization of artificial gauge fields on
neutral atomic Bose-Einstein condensates (BEC)
\cite{Lin2009_PRL,Lin2009_Nature,Lin2011_NatPhy,Lin2011}
represents one of the most important achievements in the physics
of ultracold gases. 
In the last few years, the nontrivial properties of such systems have 
attracted a broad interest, resulting in a wide number of papers devoted to
this subject (see, for example, the reviews
\cite{Dalibard2010,Goldman2014,Zhai2015} and references therein).

A very relevant class of quantum gases coupled to synthetic gauge fields
is represented by spin-orbit-coupled configurations. The first experimental
implementation of spin-orbit coupling on a neutral atomic gas was
performed by the NIST team in \cite{Lin2011}, where they managed
to realize a BEC with equal Rashba \cite{Bychkov1984} 
and Dresselhaus \cite{Dresselhaus1955} spin-orbit couplings.
The phase diagram of this system exhibits novel quantum phases,
which include a stripe phase and a spin-polarized plane-wave phase
\cite{Lin2011,Ho2011,Li2012_PRL}.
The stripe phase is characterized by periodic modulations of the density profile
resulting from the spontaneous breaking of translational symmetry, similar to
what happens in supersolids \cite{Boninsegni2012}.
Although experiments have already been made
in the relevant range of parameters, a direct evidence of such modulations is
still lacking, mainly due to the smallness of the amplitude and of the wavelength
of the fringes.

In the first part of this paper we review the ground-state properties of a
spin-orbit-coupled BEC in uniform matter, and we study how such properties
are affected by the introduction of a non-zero magnetic detuning and of
spin-asymmetric interaction strengths.
We then discuss a combined procedure to make the stripes visible and stable
\cite{Martone2014}, thus allowing for a direct experimental detection.

\section{The model and the quantum phases}
\label{sec:model_phase}
\subsection{Single-particle Hamiltonian}
\label{subsec:single_particle}
We consider a spin-$1/2$ BEC with the kind of spin-orbit-coupling first
realized by the NIST team in \cite{Lin2011}. When written in a locally
spin-rotated frame \cite{Martone2012}, the single-particle Hamiltonian
describing the system is given by (we set $\hbar=m=1$)
\begin{equation}
h_\mathrm{sp} = 
		\frac{1}{2}[\left(p_x - k_0 \sigma_z\right)^2 + p_\perp^2] 
		+ \frac{\Omega}{2} \, \sigma_x + \frac{\delta}{2} \, \sigma_z \, .
\label{eq:h_sp}
\end{equation}
It accounts for the application of two counterpropagating
and polarized laser fields, with wave vector difference $\vec{k}_0$
chosen along the $x$ direction, in the presence of a nonlinear Zeeman field.
The two lasers provide Raman transitions between the two spin states,
with Raman coupling strength given by $\Omega$. The effective magnetic
field $\delta$ is given by the sum of the true external magnetic field and
of the frequency detuning between the two lasers (see, for example,
\cite{Martone2012}). The spin matrices entering the single-particle
Hamiltonian (\ref{eq:h_sp}) are the usual $2\times 2$ Pauli matrices.
It is worth pointing out that the operator $\vec{p}$
entering (\ref{eq:h_sp}) is the canonical momentum $-i\nabla$, with
the physical velocity being given by $\vec{v}_\pm= \vec{p} \mp k_0
\hat{\vec{e}}_x$ for the spin-up and spin-down particles. In
terms of $\vec{p}$ the eigenvalues of (\ref{eq:h_sp}) are given
by
\begin{equation}
\varepsilon_{\pm}(\vec{p}) = \frac{p_x^2 + p_\perp^2}{2} + E_r \pm
\sqrt{\left(k_0 p_x - \frac{\delta}{2}\right)^2 + \frac{\Omega^2 }{4}}
\label{eq:E_single}
\end{equation}
where $E_r = k_0^2/2$ is the recoil energy. The double-branch structure
exhibited by the dispersion (\ref{eq:E_single}) reflects the spinor nature
of the system. A peculiar feature of the dispersion (\ref{eq:E_single}),
when $\delta=0$, is that it displays, as a function of $p_x$, two
degenerate minima at $\pm k_0 \sqrt{1 - (\Omega/4 E_r)^2}$, both
capable to host Bose-Einstein condensation. Notice that the wave vectors
corresponding to such minima differ from $\pm k_0$ if $\Omega \ne 0$,
and vanish for $\Omega=4 E_r$.
For larger values of $\Omega$ the gas is in the single-minimum phase,
where all the atoms occupy the $\vec {p}=0$ single-particle state.

\subsection{Many-body ground state}
\label{subsec:many_body_gs}
The peculiar features of the single-particle dispersion (\ref{eq:E_single})
are at the origin of new interesting phases in the many-body ground state
of the BEC. For a gas of $N$ particles enclosed in a volume $V$, in the
presence of two-body interactions, the mean-field interaction Hamiltonian
takes the form
\begin{equation}
H_\mathrm{int} = \int \mathrm{d}^3r 
\left[\frac{g_{\uparrow\uparrow}}{2} n_\uparrow(\vec{r})^2
+ \frac{g_{\downarrow\downarrow}}{2} n_\downarrow(\vec{r})^2
+ g_{\uparrow\downarrow}
n_\uparrow(\vec{r}) n_\downarrow(\vec{r})\right] \, ,
\label{eq:H_int}
\end{equation}
where $g_{\sigma\sigma'}=4\pi a_{\sigma\sigma'}$ ($\sigma,\sigma' = \,
\uparrow,\downarrow$) are the coupling constants in the different spin
channels, fixed by the corresponding scattering lengths $a_{\sigma\sigma'}$,
while $n_{\uparrow,\downarrow}$ are the densities of the two spin components.
The quantum phases predicted by mean-field theory depend on the value of
the relevant parameters $k_0$, $\Omega$, $\delta$ and the interaction
parameters $G_1 = \bar{n}\left(g_{\uparrow\uparrow}
+g_{\downarrow\downarrow} +2 g_{\uparrow\downarrow}\right)/8$,
$G_2 = \bar{n}\left(g_{\uparrow\uparrow}+g_{\downarrow\downarrow}
-2 g_{\uparrow\downarrow}\right)/8$, $G_3=\bar{n}\left(g_{\uparrow\uparrow}
-g_{\downarrow\downarrow}\right)/4$ \cite{Li2012_PRL},
with $\bar{n}=N/V$ the average density. In uniform matter,
the ground-state wave function can be determined through a variational
procedure based on the ansatz \cite{Li2012_PRL,Li2012_EPL,Zheng2013}
\begin{equation}
\Psi(\vec{r}) = \sqrt{\bar{n}}\left[
C_+\begin{pmatrix} \cos \theta_+ \\ -\sin\theta_+ \end{pmatrix}
e^{i k_+ x} + 
C_-\begin{pmatrix} \sin \theta_- \\ -\cos\theta_- \end{pmatrix}
e^{-i k_- x} \right] \, ,
\label{eq:ansatz}
\end{equation}
where $C_+$ and $C_-$ are coefficients
satisfying the normalization constraint $\left|C_+\right|^2+\left|C_-\right|^2=1$,
and $k_\pm$ represent the momenta at which Bose-Einstein condensation
takes place. For a given value of $k_0$, $\Omega$, $\delta$ and the two-body
interaction strengths, the values of the variational parameters $C_\pm$, $k_\pm$
and $\theta_\pm$ can be calculated through a procedure of energy minimization,
including both the single-particle (\ref{eq:h_sp}) and the interaction (\ref{eq:H_int})
terms in the Hamiltonian.
In particular, energy minimization with respect to $k_\pm$ yields the general
relationship $2\theta_\pm = \arccos\left(k_\pm/k_0\right)$ fixed by the
single-particle Hamiltonian (\ref{eq:h_sp}). Once all the variational parameters
have been determined, one can calculate key physical quantities like, for
example, the momentum distribution, accounted for by the parameters $k_\pm$,
the densities $n_\uparrow$ and $n_\downarrow$ of the two spin components,
the total density $n=\Psi^\dagger\Psi=n_\uparrow + n_\downarrow$, the
spin densities $s_k=\Psi^\dagger\sigma_k\Psi$ with $k=x,\,y,\,z$
and the corresponding spin polarizations $\langle\sigma_k\rangle =
N^{-1}\int \mathrm{d}^3 r \, s_k$.

\begin{figure}[t]
\centering
\includegraphics{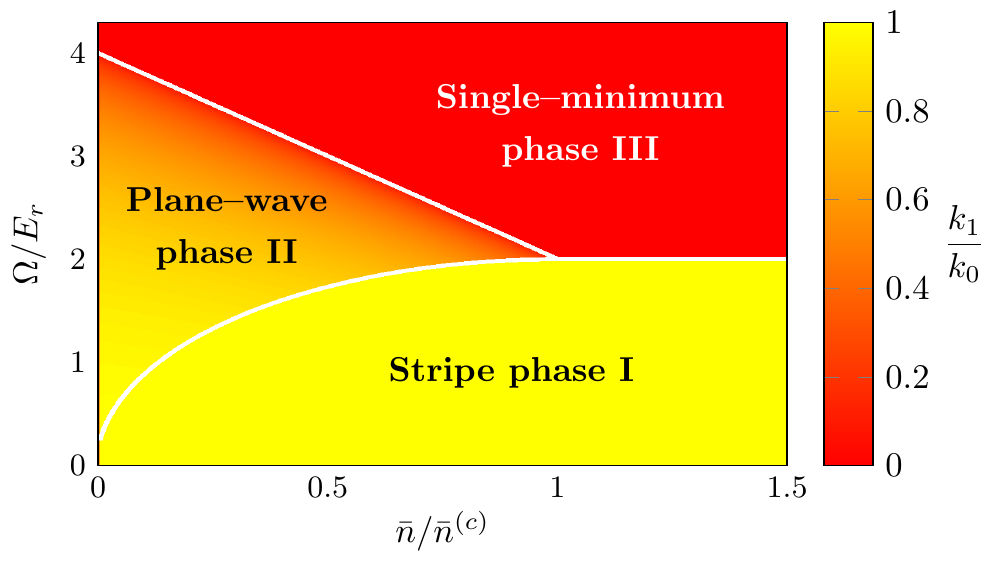}
\caption{Phase diagram of a spin-orbit-coupled BEC. The color
represents the value of $k_1/k_0$. The white solid lines identify
the phase transitions. The quantity $\bar{n}^{(c)}=E_r/(g\gamma)$
is the density at the tricritical point where
the three phases meet. The diagram corresponds to a configuration
with $\gamma = (g - g_{\uparrow\downarrow}) /
(g + g_{\uparrow\downarrow}) = 0.0012$
consistent with the value of \cite{Lin2011}.
}
\label{fig:phase_diagram}
\end{figure}

The full variational calculation has been performed in \cite{Li2012_PRL},
where the case of spin-symmetric coupling constants
$g_{\uparrow\uparrow} = g_{\downarrow\downarrow} \equiv g$
and zero detuning $\delta$ was mainly considered.
In this situation one finds $k_+ = k_- \equiv k_1$ and
$\theta_+ = \theta_- \equiv \theta$ for symmetry reasons.
The ground state was found to be compatible with three distinct quantum
phases; the corresponding phase diagram is shown
in Fig.~\ref{fig:phase_diagram}.

\vspace{1mm}

\textbf{(I) Stripe phase.} For small values of the Raman coupling
$\Omega$ and $g>g_{\uparrow\downarrow}$, the ground state is a
coherent superposition of the two plane-wave states $e^{\pm i k_1 x}$
with equal weights ($|C_+|=|C_-|=1/\sqrt{2}$), yielding a
vanishing longitudinal spin polarization $\left\langle \sigma_z\right\rangle$.
The most striking feature of this phase is the appearance of density
modulations in the form of stripes according to the law
\begin{equation}
n({\bf r})=\bar{n}\left[1+\frac{\Omega}{2\left(2E_r+G_1 \right)}
\cos\left(2k_1 x+\phi\right) \right] \, ,
\label{eq:density_stripe}
\end{equation}
with the periodicity of the fringes $\pi/k_1$ fixed by the wave vector
$k_1=k_0\sqrt{1-[\Omega/(2(2E_r+ G_1))]^2}$.
These modulations appear as the result of a mechanism of spontaneous
breaking  of translational invariance, with the actual
position of the fringes being given by the value of the phase $\phi$.
The contrast in $n({\bf r})$ is given by
\begin{equation}
\frac{n_{\rm max}-n_{\rm min}}{n_{\rm max}+n_{\rm min}}
=\frac{\Omega}{2(2E_r+G_1)}
\label{eq:contrast}
\end{equation}
and vanishes as $\Omega \to 0$ as a
consequence of the orthogonality of the two spin states entering
Eq.~(\ref{eq:ansatz}) (in this limit $\theta \to 0$ and $k_1
\to k_0$). It is worth mentioning that the ansatz,
Eq.~(\ref{eq:ansatz}), for the stripe phase provides only a first
approximation, which ignores higher-order harmonics caused
by the nonlinear interaction terms in the Hamiltonian.

\textbf{(II) Plane-wave phase.} For larger values of the Raman
coupling, the system enters the so-called plane-wave
phase (also called the spin-polarized or de-mixed phase), where
Bose-Einstein condensation takes place
in a single plane-wave state with momentum $\mathbf{p}=k_1
\hat{\mathbf{e}}_x$ ($C_- = 0$), lying on the $x$ direction (in the
following we choose $k_1>0$). In this phase, the density is uniform
and the spin polarization is given by $\langle \sigma_z\rangle =
k_1/k_0$ with
$k_1=k_0\sqrt{1-[\Omega/(4(E_r-G_2))]^2}$.
An energetically equivalent configuration is obtained by considering
the BEC in the single-particle state with
$\mathbf{p}=-k_1\hat{\mathbf{e}}_x$ ($C_+ = 0$). The choice between the two
configurations is determined by a mechanism of spontaneous symmetry
breaking, typical of ferromagnetic configurations.

\textbf{(III) Single-minimum phase.} At even larger values of
$\Omega$, the system enters the single-minimum phase (also called
zero-momentum phase), where the condensate has zero momentum ($k_1=
0$), the density is uniform, and the average spin polarization
$\langle \sigma_z\rangle$ identically vanishes, while $\langle
\sigma_x \rangle=-1$.

The critical values of the Rabi frequencies $\Omega$
characterizing the various phase transitions can be identified
by imposing that the chemical potential $\mu(\bar{n})$ 
and the pressure $P=\bar{n}\mu(\bar{n})-\int \mu(\bar{n})\,
\mathrm{d}\bar{n}$ be equal in the two phases at equilibrium.
The transition between the stripe and
the plane-wave phases has a first-order nature.
In the low density (or weak coupling) limit, i. e. $G_1,\,G_2 \ll E_r$,
the critical value of the Raman coupling $\Omega^{({\rm I-II})}$
characterizing such transition is given by the density-independent expression
\cite{Ho2011,Li2012_PRL}
\begin{equation}
\Omega^{({\rm I-II})} = 4 E_r \sqrt{\frac{2\gamma}{1+2\gamma}} \, ,
\label{eq:OmegaI-II}
\end{equation}
with $\gamma=G_2/G_1$. The transition between the plane-wave
and the single-minimum phases has instead a second-order nature
and takes place at the higher value \cite{Li2012_PRL}
$\Omega^{({\rm II-III})}=4\left(E_r - G_2\right)$,
provided that the condition $\bar{n} < \bar{n}^{(c)}$ is satisfied.
For higher densities one has instead a first-order
transition directly between the stripe and the single-minimum phases.
We also remark that, if $g<g_{\uparrow\downarrow}$, the stripe phase is 
energetically unfavorable, and only the plane-wave and the single-minimum
phases are available.

\subsection{Effects of non-zero detuning and spin-asymmetric interactions}
\label{subsec:spin_asym}
The results discussed in the previous paragraph can be easily generalized
to account for the presence of a non-vanishing magnetic detuning $\delta$
and of spin-asymmetric interactions $g_{\uparrow\uparrow} \neq
g_{\downarrow\downarrow}$. In particular, the effects of an asymmetry
in the intraspecies coupling constants can be compensated by choosing
a magnetic detuning $\delta = -2 G_3$, which ensures that the ground-state
properties remain the same as in the symmetric case \cite{Li2012_PRL}.
This holds for any value of $G_3$ in the plane-wave and the single-minimum
phases; instead, in the stripe phase exact compensation is possible only
if $G_3$ is small.

\begin{figure}
\centering
\includegraphics{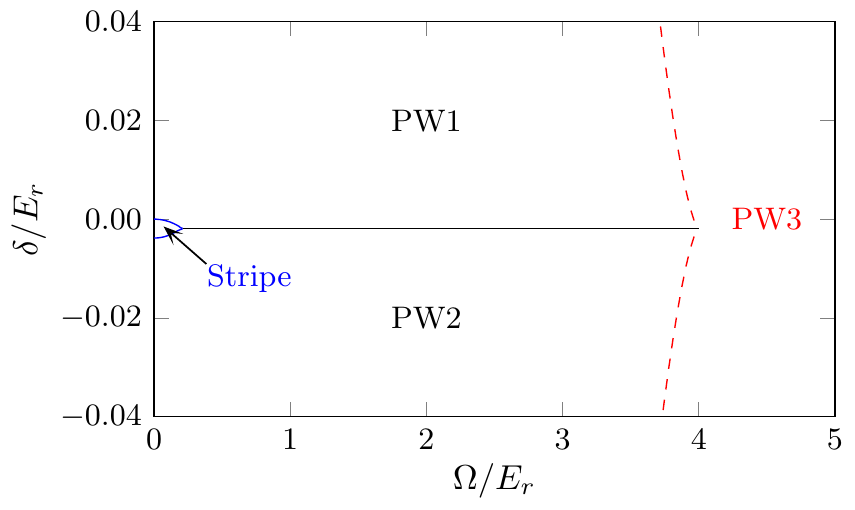}
\caption{Detuning versus Rabi coupling phase diagram in the experimental
conditions of \cite{Lin2011}. The parameters are $E_r = 2\pi\times 1.77\,$kHz,
density in the center of the trap $n_0 = 1.9\times 10^{14}$cm$^{-3}$ and
the scattering lengths given in the main text.}
\label{fig:Omega_delta_std}
\end{figure}

In the most general case of arbitrary $\delta$ and $G_3$, the ground-state
wave function can be still worked out by resorting to the ansatz
(\ref{eq:ansatz}), where now the two wave vectors $k_+$ and $k_-$
can have different values. The resulting phase diagram in the
$\Omega$-$\delta$ plane is shown in Fig.~\ref{fig:Omega_delta_std}. 
It is characterized by the occurrence
of a stripe phase, where both momentum components
of (\ref{eq:ansatz}) are present (although they can have different
weights $\left|C_+\right|$ and $\left|C_-\right|$), and by
several plane-wave states, having different values of the momentum
and hence of the magnetization \cite{Zheng2013}.

The stripe phase occurs only in configurations where $g_{\uparrow\uparrow}
g_{\downarrow\downarrow} > g_{\uparrow\downarrow}^2$,  corresponding
to the condition of miscibility of the two spin components in the absence of
spin-orbit and Raman coupling. One can notice that, in systems with almost equal
coupling constants, it occupies a very small region in the $\Omega$-$\delta$ plane.
For example, in the case of the states $\left|\uparrow\right\rangle =
\left| F=1,m_F=0\right\rangle$ and $\left|\downarrow\right\rangle =
\left| F=1,m_F=-1\right\rangle$ of $^{87}$Rb, where $a_{\uparrow\uparrow}=
101.41\,a_B$ and $a_{\downarrow\downarrow}=a_{\uparrow\downarrow}=
100.94\,a_B$, from  Eq.~(\ref{eq:OmegaI-II}) one finds the value
$\Omega^{({\rm I-II})} = 0.19 \, E_r$ for the critical Rabi coupling,
while the critical magnetic detuning needed to bring the system from the
stripe to the spin-polarized phase turns out to be of the order of $10^{-3}\, E_r$.
The latter is actually proportional to the difference $\Delta\mu$
between the chemical potentials in the two phases. An analytic estimate 
can be obtained in the $\Omega \to 0$ limit, where one finds that the stripe
phase is favored for values of the magnetic detuning $\delta$ such that
$\left|\delta + 2 G_3\right| \leq 4 G_2$.
At finite values of $\Omega$ the range of values of $\delta$ compatible
with the stripe phase is further reduced. As a consequence, a tiny magnetic
field (arising, for instance, from external fluctuations) can easily bring the
system into the spin-polarized phases. The stability of the stripe phase
can be strongly enhanced if one increases significantly the value of $G_2$,
as we will discuss in Sect.~\ref{sec:exp_stripes}.

The phase diagram of Fig.~\ref{fig:Omega_delta_std} also contains three different
kinds of plane-wave phase. Those on the left region of the diagram, denoted by
PW1 and PW2, correspond to the regime where the single-particle Hamiltonian
has two local minima; PW1 and PW2 are favored for magnetic detunings $\delta$
larger and smaller than $-2G_3$, respectively. The state PW3 appears instead
in the region where the single-particle Hamiltonian has one local minimum only.
The red dashed lines represent the transition between the double-minimum
and the single-minimum regimes; in drawing them we have taken the presence of
spin-dependent interactions into account, yielding some corrections with
respect to the single-particle results \cite{Zheng2013}.

\section{Experimental perspectives for the stripe phase}
\label{sec:exp_stripes}
The stripe phase is doubtlessly the most intriguing phase appearing
in the phase diagram of Sect.~\ref{sec:model_phase}. It has been
the object of several recent theoretical investigations
\cite{Ho2011,Wang2010,Wu2011,Sinha2011,Ozawa2012,Li2013,Zezyulin2013,Lan2014,Sun2014,Han2014,Hickey2014}.
The stripe phase is characterized by the spontaneous breaking of
two continuous symmetries. The breaking of gauge symmetry yields
superfluidity, while the breaking of translational invariance is
responsible for the occurrence of a crystalline structure. The
simultaneous presence of these two broken symmetries is typical of
supersolids
\cite{Boninsegni2012,Andreev1969,Leggett1970,Chester1970}.
It has been shown, among other things, to be at the origin
of the appearance of two gapless excitations as well as of a band
structure in the excitation spectrum \cite{Li2013}.

In the experiments of \cite{Lin2011} and \cite{Ji2014} a phase
transition has been detected close to the theoretical prediction
$\Omega^{({\rm I-II})} = 0.19 \, E_r$ (see Eq.~(\ref{eq:OmegaI-II}))
for the critical Raman coupling below which the occurrence of the stripe
phase is expected. However, as we already mentioned in the introduction,
there is still no direct experimental evidence of the periodic 
modulations in the density profile characterizing the stripe phase.
The main reason is that, in the conditions of current experiments 
with spin-orbit-coupled $^{87}$Rb BECs 
\cite{Lin2011,Zhang2012,Khamehchi2014},
the contrast and the wavelength of the fringes are too small
to be revealed. Another issue is represented by the fragility
of the stripe phase against fluctuations of external magnetic
fields, which has already been discussed in
Par.~\ref{subsec:spin_asym}. In \cite{Martone2014} the authors
proposed a procedure to make
the experimental detection of the fringes a realistic perspective,
improving their contrast and their wavelength, and increasing
the stability of the stripe phase against magnetic fluctuations.

In order to achieve a larger value of the contrast
(\ref{eq:contrast}), one needs to enlarge the range of values
of $\Omega$ compatible with the existence of the stripe
phase. As can be seen from Eq.~(\ref{eq:OmegaI-II}),
an efficient way to increase the critical Raman coupling
$\Omega^{({\rm I-II})}$ is to reduce the value of the interspecies
coupling constant $g_{\uparrow\downarrow}$. A possibility is to look
for hyperfine states characterized by a small (or tunable)
interspecies scattering length.
Here we discuss a different strategy, based on the idea of reducing
the effective interspecies coupling by means of suitable trapping
conditions. In particular, one can trap the atomic gas in a 2D
configuration, with tight confinement of the spin-up and spin-down
components around two different positions, displaced by a distance
$d$ along the $z$ direction.
This configuration can be realized with a trapping potential
of the form  
\begin{equation}
V_\mathrm{ext}(z) = 
\frac{\omega^2_z}{2}\left(z-\frac{d}{2}\sigma_z\right)^2
\label{eq:Vextd}
\end{equation}
produced either through magnetic gradient techniques or
via spin-dependent optical potentials.
Assuming a Gaussian profile $\psi_{\pm} =
(1/\sqrt[4]{\pi a_z^2})e^{-(z\mp d/2)^2/2a^2_z}$
for the $z$ dependence of the spin-up and spin-down wave functions,
with $a_z=1/\sqrt{\omega_z}$ the oscillator length along $z$,
the integration over $z$ of the energy functional (\ref{eq:H_int})
gives rise to effective 2D coupling constants
$\tilde{g}_{\alpha\beta}$ given by\footnote{In the present section
we consider realistic spin-asymmetric interaction strengths
$g_{\uparrow\uparrow} \neq g_{\downarrow\downarrow}$, and we
compensate the asymmetry by choosing $\delta= - 2 G_3$.}
\begin{equation}
\tilde{g}_{\uparrow\uparrow,\downarrow\downarrow}
=\frac{1}{\sqrt{2\pi}a_z} g_{\uparrow\uparrow,\downarrow\downarrow} 
\; , \qquad
\tilde{g}_{\uparrow\downarrow}
=\frac{1}{\sqrt{2\pi}a_z} g_{\uparrow\downarrow}e^{-d^2/2a^2_z} \, .
\label{eq:gtilde}
\end{equation}
In an analogous way one finds that also the effective Raman coupling,
to be used in 2D, is lowered with respect to the physical coupling
$\Omega$ according to the law $\tilde{\Omega} =  e^{-d^2/4a^2_z }\Omega$, 
reflecting the reduction of the overlap between the two wave functions.
Hence, the new configuration produced by 
a tight axial trapping potential with a spin-dependent displacement 
can be described formulating the Hamiltonian in 2D, with 
the effective Raman coupling given by $\tilde{\Omega}$,
and the interaction term obtained from the functional (\ref{eq:H_int})
with the replacement of the 3D densities with their 2D counterparts
$\int \mathrm{d}z \, n$ and of the coupling constants with the renormalized
values (\ref{eq:gtilde}). 
The main advantage with respect to the original 3D problem
is that now, due to the relative separation of the atomic clouds
of the two spin components, the effective $g_{\uparrow\downarrow}$
is reduced with respect to the two intraspecies coupling constants
(see Eq.~(\ref{eq:gtilde})). As a consequence, the ratio $\gamma$
appearing in Eq.~(\ref{eq:OmegaI-II}) is now given by the expression
\begin{equation}
\gamma = \frac{\tilde{G}_2}{\tilde{G}_1} = 
\frac{g_{\uparrow\uparrow}+g_{\downarrow\downarrow}
-2g_{\uparrow\downarrow}e^{-d^2/2a^2_z}}
{g_{\uparrow\uparrow}+g_{\downarrow\downarrow}
+2g_{\uparrow\downarrow}e^{-d^2/2a^2_z}} \, ,
\label{eq:gamma}
\end{equation}
and is larger with respect to the 3D case, yielding an increased value
of the critical effective Raman coupling,
and thus of the largest reachable contrast of the fringes in the stripe
phase. For example, choosing the value $d=a_z$ and the $^{87}$Rb
hyperfine states mentioned in Par.~\ref{subsec:spin_asym},
one finds the value $\gamma=0.25$ for the ratio (\ref{eq:gamma}),
to be compared with the value $\gamma =0.0012$
for the $d=0$ case\footnote{Another
important change is that, due to the increase of the 
value of $\gamma$, the critical density $n^{(c)}$ can be
significantly lowered with respect to the value in the $d=0$
case, becoming of more realistic achievement in future
experiments.}.

\begin{figure}
\centering
\includegraphics{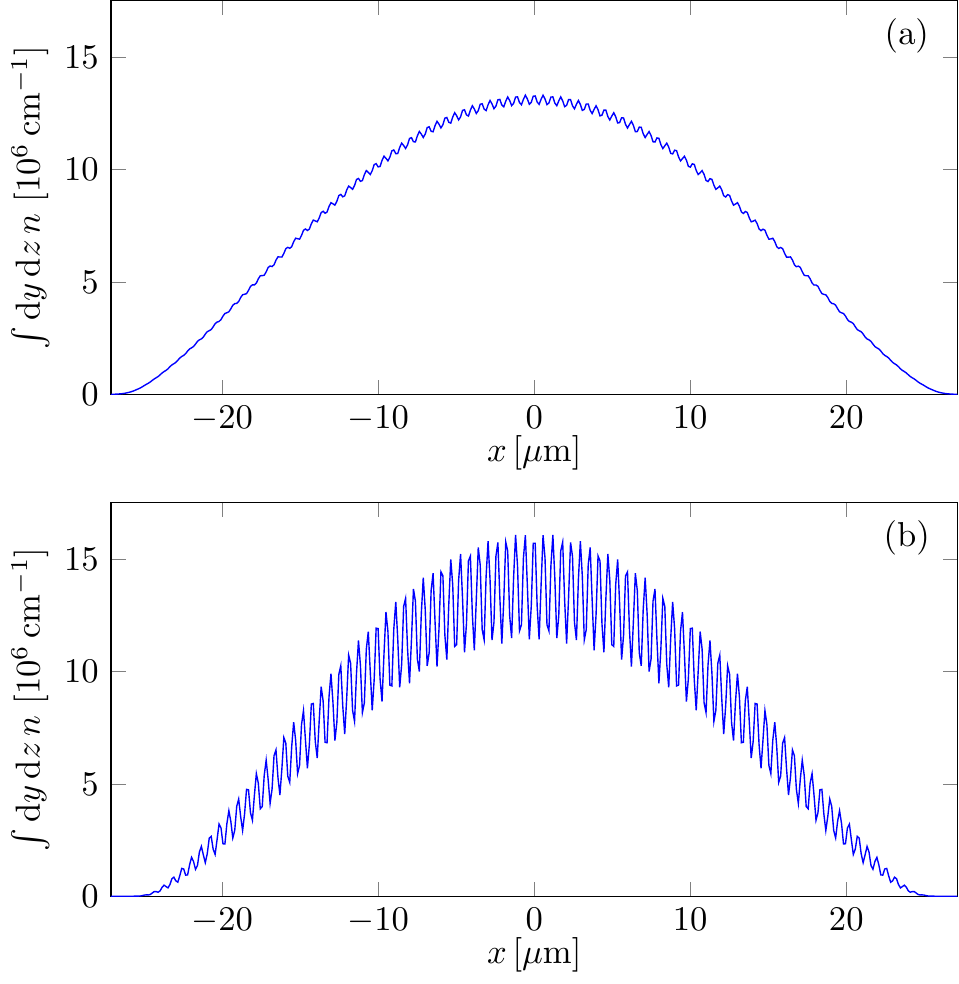}
\caption{Integrated density profile $\int {\rm d}y \, {\rm d}z \, n$ in the
striped phase. (a) displays the situation without separation of the traps
for the two spin components. (b) corresponds instead to traps separated
along $z$ by a distance $d=a_z$, which increases the contrast of the fringes.}
\label{fig:density_prof}
\end{figure}

Quantitative predictions for the novel configuration discussed above 
can be obtained by solving numerically the 3D Gross-Pitaevskii equation.
In Fig.~\ref{fig:density_prof} we show the results for a gas 
of $N=4\times 10^4$ $^{87}$Rb atoms confined by a harmonic potential 
with frequencies $\left(\omega_x,\omega_y,\omega_z\right) = 
2\pi\times\left(25,100,2500\right)\,$Hz, the scattering lengths
$a_{\sigma\sigma'}$ and the recoil energy $E_r$ equal
to those reported in Par.~\ref{subsec:spin_asym}, 
and consistent with Ref.~\cite{Lin2011}.
Fig.~\ref{fig:density_prof}a corresponds to $d=0$, 
while Fig.~ \ref{fig:density_prof}b corresponds to $d=a_z=0.22\, \mu$m.
In both Fig.~\ref{fig:density_prof}a and Fig.~\ref{fig:density_prof}b
we have chosen values of the Raman coupling equal to one half
the critical value to enter the plane-wave phase, in order to ensure
more stable conditions for the stripe phase.
In Fig.~\ref{fig:density_prof}a this corresponds to 
$\Omega = (1/2) \Omega^{({\rm I-II})}(\gamma) = 0.095\, E_r$
with $\gamma=0.0012$, while in Fig.~ \ref{fig:density_prof}b to 
$\Omega = (1/2)e^{d^2/4a^2_z } \Omega^{({\rm I-II})}(\gamma)=1.47\, E_r$ 
with $\gamma=0.25$.
The plotted density corresponds to the 1D density as a function of the most
relevant $x$ variable, obtained by integrating the full 3D density along the
$y$ and $z$ direction. The figure clearly shows that in the
conditions of almost equal coupling constants (Fig.~\ref{fig:density_prof}a)
the density modulations are very small, while their effect is strongly 
amplified in Fig.~\ref{fig:density_prof}b where the interspecies coupling
is reduced with respect to the intraspecies values by a factor $\sim 0.61$.

\begin{figure}[t]
\centering
\includegraphics{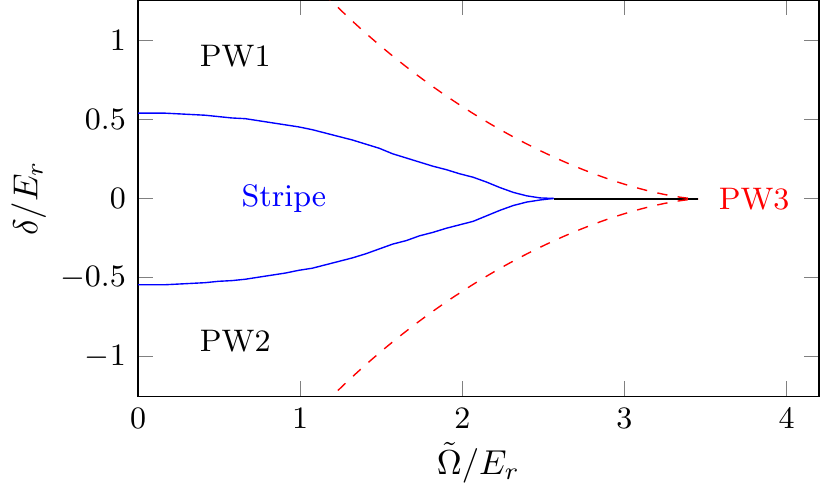}
\caption{Detuning versus effective Rabi coupling phase diagram in
the conditions of Fig.~\ref{fig:density_prof}b.}
\label{fig:Omega_delta_sim}
\end{figure}

The suggested procedure has also the positive effect of making 
the stripe phase more robust against fluctuations of external magnetic fields.
Indeed, the reduction of the interspecies coupling and the increase of the
local 3D density, due to the tight axial confinement, yield a significant
increase of the energy difference between the stripe and the plane-wave phases.
For example, in the case considered in the above 3D
Gross-Pitaevskii simulation with $d=a_z$ (Fig.~\ref{fig:density_prof}b),
a magnetic detuning of the order of $0.3 \, E_r$ is needed to bring the
system into the spin-polarized phase (see the diagram in
Fig.~\ref{fig:Omega_delta_sim}), while in the absence of displacement 
(Fig.~\ref{fig:density_prof}a) the critical value is much smaller
($\sim 0.001 \, E_r$, see Par.~\ref{subsec:spin_asym} and
Fig.~\ref{fig:Omega_delta_std}).

\begin{figure}[t]
\centering
\includegraphics{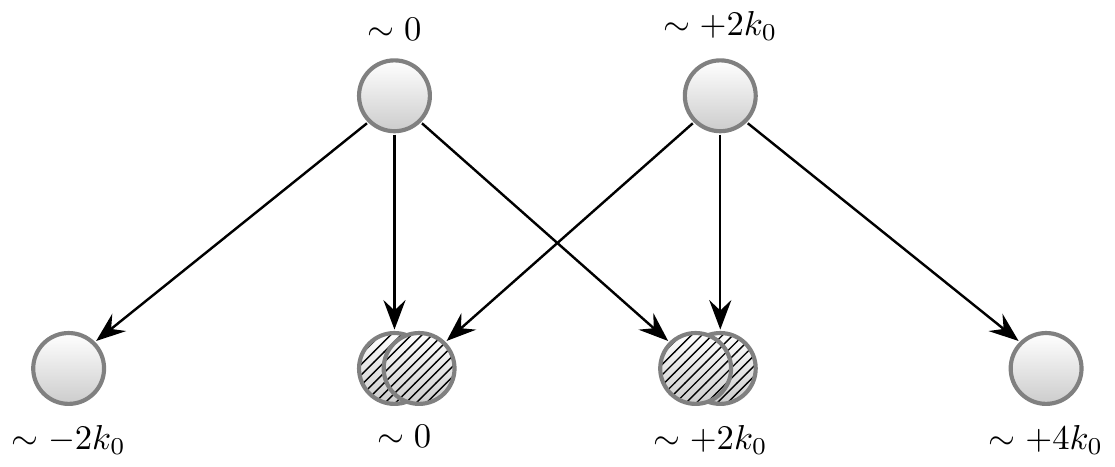}
\caption{Schematical description of the splitting of the spin-down
component of the stripe wave function into different momentum
components caused by a $\pi/2$ Bragg pulse transferring momentum
$2k_1-\epsilon$.}
\label{fig:Bragg_scheme}
\end{figure}

\begin{figure}
\centering
\includegraphics{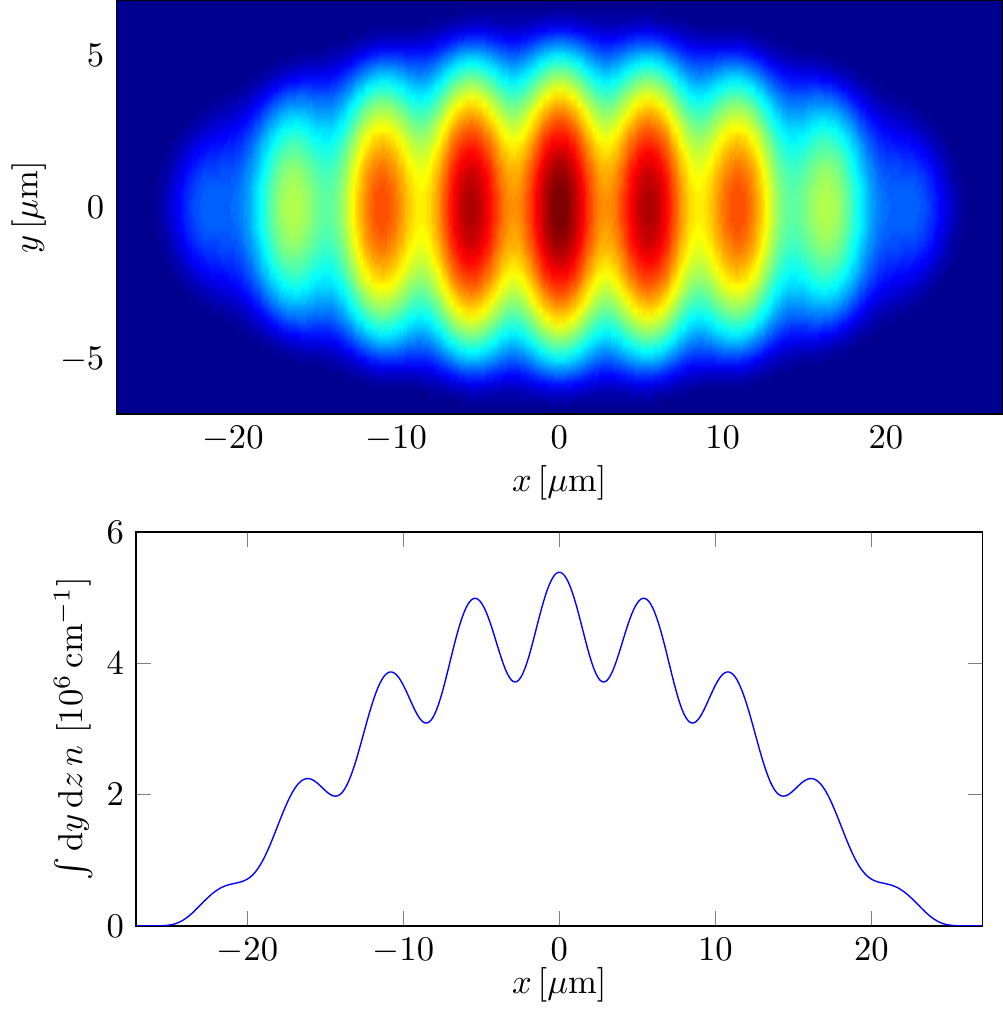}
\caption{Integrated density profiles $\int {\rm d}z \, n$ (top) and
$\int {\rm d}y \, {\rm d}z \, n$ (bottom) in the stripe phase, in the same
conditions as Fig.~\ref{fig:density_prof}b, after the application of a $\pi/2$
Bragg pulse transferring momentum $\pm 1.8\,\hbar k_1$.}
\label{fig:density_prof_Bragg}
\end{figure}

Let us finally address the problem of the small spatial separation
of the fringes, given by $\pi/k_1$, which turns out to be of the
order of a fraction of a micron in standard conditions. One
possibility to increase the wavelength of the stripes is to lower
the value of $k_0$ by using lasers with a smaller relative incident
angle. In the following we discuss a more drastic procedure which
consists of producing, after the realization of the stripe phase, a
$\pi/2$ Bragg pulse with a short time duration (smaller than the
time $1/E_r$ fixed by the recoil energy), followed by the sudden
release of the trap. This pulse can transfer to the condensate a
momentum $k_B$ or $-k_B$ along the $x$ direction, where $k_B$ is
chosen equal to $2k_1 -\epsilon$ with $\epsilon$ small compared to
$k_1$. The $\pi/2$ pulse has the effect of splitting the condensate
into various pieces, with different momenta. The situation is
schematically shown in Fig.~\ref{fig:Bragg_scheme} for the spin-down
component, where the initial condensate wave function, which in the
stripe phase is a linear combination with canonical momenta $\pm
k_1$, corresponding to momenta $k_0-k_1$ and $k_0+k_1$ in the
laboratory frame, after the Bragg pulse will be decomposed into six
pieces. Two of them, those labeled in the lower part of the figure
with momentum $\sim 0$, will be practically at rest after the pulse
and are able to interfere with fringes of wavelength
$2\pi/\epsilon$, which can easily become large and visible {\it in
situ}. It is worth noticing that these two latter pieces originate
from the two different momentum components of the order parameter
(\ref{eq:ansatz}) in the stripe phase and involve $1/3$ of the total
number of atoms. The corresponding interference effect would be
consequently absent in the plane-wave phase, where only one momentum
component characterizes the order parameter. The other pieces
produced by the Bragg pulse carry much higher momenta and will fly
away rapidly after the release of the trap and of the laser fields.
In Fig.~\ref{fig:density_prof_Bragg} we show a typical behavior of
the density profile obtained by modifying the condensate wave
function in momentum space according to the prescription discussed
above.

\section{Conclusions}
In this paper we have reviewed the ground-state properties of a
spin-orbit-coupled BEC, focusing on the effects of the presence
of a finite magnetic detuning and of spin-asymmetric coupling
constants. The phase diagram includes a stripe phase, which
is characterized by the presence of periodic modulations in the density profile.
In the last part of the paper we have discussed a combined
procedure to increase the visibility of such modulations and
the stability of the stripe phase against magnetic fluctuations,
thus favoring the exploration of this intriguing configuration
in realistic experimental conditions.

\begin{acknowledgement}
The author acknowledges stimulating discussions with Yun Li
and Sandro Stringari. This work has been supported by ERC
through the QGBE grant and by Provincia Autonoma di Trento.
\end{acknowledgement}

\end{document}